
\input phyzzx
hep-th/9410201
\def\Di{D\hskip-0.2cm /}
\def\Dir{D\hskip-0.25cm /}
\date{LMPM ../94}
\titlepage
\title{\bf LINEAR CONNECTIONS IN \break NON-COMMUTATIVE GEOMETRY}
\author{J. Mourad\foot{e-mail:MOURAD@celfi.univ-tours.fr}}
\address{Laboratoire de Mod\`eles de Physique Math\'ematique,
 Universit\'e de Tours, Parc de
Grandmont, F-37200 Tours, France}

\abstract{A construction is proposed for linear connections on
non-commutative algebras. The construction relies on a generalisation
of the Leibnitz rules of commutative geometry
and uses the bimodule structure of $\Omega^{1}$. A special role is played
by the extension to the framework of non-commutative geometry
of the permutation of two copies of $\Omega^{1}$.
The construction of the linear connection as
well as the definition of torsion and curvature
is first proposed in the setting of
the derivations based differential calculus
of Dubois-Violette and then a
generalisation to the framework
proposed by Connes
as well as other non-commutative differential
calculi is suggested. The covariant derivative
obtained admits an extension to the tensor product
of several copies of $\Omega^{1}$. These constructions are
illustrated with the example of the algebra of $n\times n$ matrices.}
\endpage

\REF\cona{A. Connes, {\it Non-Commutative Differential Geometry}, {
Publ. Math
I.H.E.S.} {\bf 62}, 257 (1985);
{ \it G\'eom\'etrie noncommutative}, (InterEditions, Paris, 1990).}
\REF\duba{M. Dubois-Violette, { C.R. Acad. Sc. Paris }
S\'erie I {\bf 297}, 403 (1988); in {\it Differential Methods in Theoretical
physics}, Proceedings Rapallo, 1990 (C. Bartocci, U. Bruzzo, R. Cianci,
eds), Lecture notes in Physics {\bf 375}, (Springer-Verlag, Berlin, Heidelberg,
New-York, 1991).}
\REF\conb{A. Connes, { Comm. Math. Phys.} {\bf 117}, 673 (1988).}
\REF\wit{E. Witten, {  Nucl. Phys.} {\bf B268}, 253 (1986).}
\REF\wie{H.W. Wiesbrock, { Comm. Math. Phys.} {\bf 136}, 369 (1991).}
\REF\for{P. Forgacs, N.S. Manton, { Comm. Math. Phys.} {\bf 72}, 15 (1980).}
\REF\dubb{M. Dubois-Violette, R. Kerner, J. Madore, { Phys. Lett.}
{\bf B217}, 485
(1989); { Class. Quant. Grav.} {\bf 6}, 1709 (1989);
{ J. Math. Phys.} {\bf 31}, 323 (1990); { Class. Quant. Grav.}
 {\bf 8}, 1077 (1991).}
\REF\dubc{M. Dubois-Violette, R. Kerner, J. Madore, { J. Math.
Phys.} {\bf 31}, 316 (1990).}
\REF\conc{A. Connes, J. Lott, { Nucl. Phys. Proc. Suppl.} {\bf B18}, 29
 (1989).}
\REF\cond{A. Connes, {\it Noncommutative Geometry} (book in press,
Dover Publications).}
\REF\kas{D. Kastler, Rev. Math. Phys. {\bf 5}, 477 (1993).}
\REF\coq{R. Coquereaux, G. Esposito-Far\`ese, G. Vaillant, {Nucl.
Phys.} {\bf B353}, 689 (1991).}
\REF\Bal{B.S. Balakrishna, F. G\"ursey, K.C. Wali, { Phys. Rev.} {\bf D44},
3313 (1991).}
\REF\chama{A.H. Chamseddine, G. Felder, J. Fr\"ohlich, {Phys.
Lett.} {\bf B296}, 109 (1992); {Nucl. Phys.} {\bf B395}, 672 (1993).}
\REF\var{J.C. Varilly, J.M. Gracia-Bondia, { Jour. of Geom.
and Phys.} {\bf 12}, 223 (1993).}
\REF\alv{E. Alvarez, J.M. Gracia-Bondia, C.P. Martin
 {Phys. Lett.} {\bf B306}, 55 (1993); { Phys. Lett} {\bf B329}, 259
 (1994).}
\REF\mada{J. Madore, { Phys. Rev.} {\bf D41}, 3709 (1990).}
\REF\madb{J. Madore, J. Mourad, {  Class. Quant. Grav.} {\bf 10}, 2157
 (1993);
{ Int. J. Mod. Phys.} {\bf D3}, 221 (1994).}
\REF\chamb{A.H. Chamseddine, G. Felder, J. Fr\"ohlich, { Comm.
Math. Phys.} {\bf 155}, 205 (1993).}
\REF\sit{A. Sitarz, {\it Gravity from noncommutative geometry}, { preprint}
 TPJU 1-94 (1994), hep-th@xxx.lanl.gov/9401145.}
\REF\dubd{M. Dubois-Violette, P.W. Michor, {\it Derivations
and noncommutative differential calculus 2}, { preprint LPTHE-Orsay
 94/50,}
hep-th@xxx.lanl.gov/9406166; {\it The Fr\"olicher-Nijenhuis bracket
for derivation based non commutative differential forms},
{preprint} dg-ga@msri.org/9410007.}
\REF\kal{W. Kalau, M. Walze, {\it Gravity, non-commutative geometry
and the Wodzicki residue}, preprint MZ-TH 93/38 (Dec. 1993),
gr-qc@xxx.lanl.gov/9312031.}
\REF\wes{J. Wess, B. Zumino, Nucl. Phys. Proc. Suppl. {\bf 18B},
302 (1990).}
\REF\wor{S.L. Woronowicz, Comm. Math. Phys. {\bf 111}, 613 (1987);
{\it ibid.} {\bf 122}, 125 (1988).}
\chapter{Introduction}

Non-commutative geometry [\cona,\duba] offers a novel and promising
framework for the construction of physical theories.
The basic idea is to replace the commutative algebra
of functions on a manifold by a more general associative
algebra. Geometrical objects such as forms and  gauge fields
are constructed using the algebraic structure which replaces
the manifold structure of commutative geometry. The exterior
differential calculus and vector bundles have been succesfully
generalised to the non-commutative context.
An analog of integration has also been proposed by Connes [\conb],
it uses the Dixmier trace and the Wodzicki residue.

Non-commutative geometry has been first used for physics by Witten
[\wit,\wie]
as a suitable framework for open string field theory where
the exterior derivative is naturally provided by the BRST
charge. The action used by Witten is a topological one,
it does not make use of a metric or a linear connection.
In the spirit of gauge theories on higher dimensions [\for],
gauge theories were constructed on algebras of the form
${\cal C}(M)\otimes {\cal A}$, where ${\cal C}(M)$
is the algebra of functions on
four-dimensional space-time and ${\cal A}$ is a discrete
algebra [\dubb-\var].
These theories lead to a natural
appearance of the Higgs field with a symmetry breaking potential.
The standard model with some constraints on its
parameters has been succesefully constructed
using this strategy [\conc,\cond,\kas].
 These constraints cannot, however, be implemented in
a renormalization group invariant way [\alv].
In the same spirit, gravitational theories have been constructed
on the same type of algebras [\mada,\chamb,\madb,\sit]. In the particular
case where the algebra ${\cal A}$ is that of
$n\times n$ matrices the resulting
theory is the truncated version of Kaluza-Klein theory with
the group
$SU_{n}$ as an internal manifold [\madb].
These constructions relied heavily
on the generalisation of Cartan's structure equations. They
used a moving basis to define the connection form.
A proposal for the construction
of a linear connection on ${\cal C}(M)\otimes Z_{2}$
independently of the metric
has been done [\chamb,\sit] using an axiomatic generalisation
of the Leibnitz rule. This rule uses the left module structure
of $\Omega^{1}$.

The purpose of this article is to define linear connections
on general non-commutative algebras. Our construction
uses the bimodule structure of $\Omega^{1}$. That is
we postulate {\it two} Leibnitz rules that allow to
calculate the covariant derivative of $f \omega$
and $\omega f$, where $f$ is an element of $\cal A$
and $\omega$ an element of $\Omega^{1}$, from the knowledge of
the covariant derivative of $\omega$. The situation is thus
as in commutative geometry. This construction makes possible
the extension of the covariant derivative
to the tensor product over $\cal A$ of several copies of $\Omega^{1}$.
Had we used only the left module structure of $\Omega^{1}$
this extension could not be possible. Needless to mention, the
importance of the linear connection for the formulation of
gravitational theories on general non-commutative algebras
cannot be overemphasised (see however [\kal]).

Section 2 fixes the notation and
is a brief review of non-commutative differential
calculi, that based on derivations proposed by Dubois-Violette and
the one pioneered by Connes.
 In section 3, we review the
construction of  linear connections in commutative geometry using
a formulation that will be appropriate for
a non-commutative generalisation which is the subject of
section 4. In this section, a definition for a covariant
derivative is proposed for the differential calculus
of Dubois-Violette. This definition suggests
a way of constructing linear connections in
the setting of other differential calculi. The construction
of linear connections is reduced to the generalisation
to the non-commutative setting of a permutation acting
on the tensor product of two copies of $\Omega^{1}$.
Section 4 ends with the study of linear connections
on matrix algebras.
We collect our conclusions in section 5.

\chapter{Non-commutative differential calculi}

\section{The differential calculus of Dubois-Violette}

Let $\cal A$ be an associative *algebra with unity and {\cal D}er
the set of derivations of ${\cal A}$. An element $X$ of
${\cal D}$er is a linear map from ${\cal A}$ to itself satisfying
the Leibnitz rule:
$$X(fg)=X(f)g+fX(g) \eqn\le,$$
where $f$ and $g$ are elements of $\cal A$. ${\cal D}er$ offers
a generalisation of the Lie algebra of vector fields over a manifold.
Contrary to the commutative case, $\cal D$er does not have
the structure of an $\cal A$-module, that is if $X$ is
a derivation $fX$ does not represent a derivation unless
$f$ is in the center of $\cal A$.
A 1-form is defined as a linear map from $\cal D$er to $\cal A$,
the set of 1-forms is denoted by $\underline{\Omega}^{1}_{{\cal D}\rm er}$.
A p-form is defined to be a skew symmetric multilinear map over $Z({\cal
A})$, the center of $\cal A$,
from $\overbrace{{\cal D} {\rm er}\otimes_{Z({\cal A})}
 \dots \otimes_{Z({\cal A})}{\cal D} {\rm er}}^
{\rm  p \ times}$ to $\cal A$.
The wedge product may be defined as in commutative geometry
by appropriate antisymmetrisation (for details see [\duba]).
For example,
the wedge product of two 1-forms $\omega$ and
$\omega'$ is given by:
$$\omega\wedge\omega' (X_{1},X_{2})=
\omega(X_{1})\omega'(X_{2})-\omega(X_{2})\omega'(X_{2}).\eqn\pro$$
Note that, in general, $\omega\wedge\omega'$
and $\omega'\wedge\omega$ are not simply related.

The exterior derivative of a p-form is a p+1-form defined,
as in commutative geometry, by the formula:
$$\eqalign{
d\omega (X_{1},\dots,X_{p+1})= &\sum_{i}
X_{i}(\omega(X_{1},\dots,X_{i-1},\hat X_{i},X_{i+1},\dots,X_{p+1}))\cr
&\sum_{1\leq i<j\leq p+1}
 (-1)^{i+j}\omega([X_{i},X_{j}],X_{1}, \dots,
\hat X_{i},\dots,\hat X_{j},\dots,X_{p+1}).}\eqn\dex$$
Here, the notatation $\hat X$ means that $X$ is omitted.
The exterior derivative verifies $d^{2}=0$ and the graded Leibnitz
rule:
$$d(\omega \wedge \omega')=d\omega\wedge\omega'+
(-1)^{\rm deg \omega}\omega\wedge d\omega'.\eqn\grle$$
The set $\underline{\Omega}_{{\cal D}\rm er}
=\oplus\underline{\Omega}^{p}_{{\cal D}\rm er}$
is thus a graded differential algebra.
In fact, Dubois-Violette
considers two graded differential algebras [\duba]:
the other one, $\Omega_{{\cal D}er}$
is the smallest graded differential algebra
in $\underline{\Omega}_{{\cal D}er}$
containing ${\cal A}$. This distinction will not be important
in the following. For further developments concerning
this derivation based differential calculus see [\dubd].

\section{The differential calculus of Connes}

The differential calculus of Connes [\cona,\cond] does not refer
to derivations for the construction of differntial forms.
In one formulation it relies on a BRST like charge
and the exterior derivative is a graded commutator, and in
another more general formulation, it uses an analog of the Dirac
operator.

\subsection{The exterior derivative as a graded commutator}

The starting point is the imbedding of the
algebra $\cal A$ in the algebra $l({\cal H})$
of operators on a graded Hilbert space ${\cal H}$.
An odd operator, $Q$, is chosen in $l({\cal H})$
such that its square is in the center of $l({\cal H})$.
The exterior derivative is taken to be the graded commutator with $Q$.
It satisfies, as it should, the graded Leibnitz
rule as well as the relation $d^{2}=0$. The resulting
graded differential algebra is denoted by $\Omega^{*}_{Q}$.

Note that the differential calculus used by Witten
relies on this construction. In his work,
 grading is provided by the ghost number and $Q$
is taken to be the BRST charge.


\subsection{Differential forms with Dirac operators}

The method consists in taking the quotient
of the universal differential algebra, $\Omega^{*}$,
by some differential  ideal $J_{\Di}$. The differential
ideal is stable under left and right multiplication by elements
of $\cal A$ and under the universal exterior derivative, $\delta$.
The construction of $J_{\Di}$ uses the following
ingredients:

[1] A graded Hilbert space $\cal H$ such
that there exists a representation, $\rho$, of the
algebra $\cal A$ on the algebra of operators on $\cal H$,
$l({\cal H})$.

[2] An odd unbounded operator $\Dir$ in $l({\cal H})$ such that
$[\Dir,\rho(f)]$ is bounded for all elements
$f$ of $\cal A$. In the commutative case the
operator $\Dir$ is taken to be the Dirac operator
on the manifold.

Consider the extension of $\rho$ on $\Omega^{*}$
defined by
$$\rho^{*}(f_{0} \delta f_{1}\dots \delta f_{p})=\rho(f_{o})
[\Dir,\rho(f_{1})]\dots [\Dir,\rho(f_{p})].$$
Let $J$ be the kernel of  $\rho^{*}$
and define the graded ideal $J_{gr}$ by $\oplus_{n}J\bigcap\Omega^{n}$
and the graded differential
ideal $J_{\Di}$ by $J_{gr}+\delta J_{gr}$. The quotient
$\Omega^{*}_{\Di}=\Omega^{*} /J_{\Di}$ is the differential algebra
advocated by Connes to generalise the de Rahm algebra on a
compact Euclidien spin manifold. Note that one can
identify $\Omega_{\Di}^p$ with $\rho^*(\Omega^p)/(\rho^{*}
 (\delta J_{gr}^{p-1}))$.
Suppose the $n^{\rm th}$ eigenvalue of the operator
$\Dir$ grows as $n^{1 \over d}$ for $n \rightarrow \infty$. One may
define a scalar product in the Hilbert space completion of
$\rho^{*}(\Omega^{p})$ [\cond]:
$$<A,B>=\Tr_\omega(A^{*}B|\Di|^{-d})=\lim_{t \rightarrow \infty}
t^{d \over 2}\Tr (A^{*}B {\rm e}^{-t{\Di}^{2}}).\eqn\sca$$
The Hilbert space completion of $\Omega_{\Di}^{p}$ may then be
identified with  the orthogonal to $\rho^{*}(\delta J_{gr}^{p-1})$.

Note that this construction encodes more
information than the exterior differential calculus
since the Dirac operator on a manifold depends on the metrical
properties of the manifold.

\chapter{Linear connections in commutative geometry}

In this section the construction of the linear connection
of commutative geometry is done in a
manner which we will find appropriate
for a non-commutative generalisation. In this section,
we denote by $\Omega^{p}$
the set of de Rahm p-forms.

The covariant derivative is a linear map
from $\Omega^{1}$ to $\Omega^{1}\otimes_{{\cal C}(M)}\Omega^{1}$
verifying the following Leibnitz
rule:
$$\nabla(f \omega)=df\otimes\omega+f\nabla \omega.\eqn\leia$$
Since functions commute with forms
we also have the following Leibnitz rule:
$$\nabla(\omega f)=\sigma(\omega\otimes df)+\nabla \omega f,\eqn\leie$$
where $\sigma$ is a permutation acting on $\Omega^{1}
\otimes_{{\cal C}(M)}\Omega^{1}$:
$$\sigma(\omega\otimes\omega')=\omega'\otimes\omega.\eqn\perm$$
The two Leibnitz rules \leia \ and \leie \ are equivalent.
We will see that in non-commutative geometry this will not be the case.

Suppose $M$ is parallelisable and let $b^{a}$ be a basis
of $\Omega^{1}$, then the covariant derivative of a 1-form
is uniquely determined by the covariant derivatives of the basis
$b^{a}$. The linear connection is defined by $\Gamma^{a}=
-\nabla b^{a}$.
Let  $m$ be
the multiplication map from $\Omega^{1}\otimes_{{\cal C}(M)}\Omega^{1}$
to $\Omega^{2}$.
  The
operator $T=d-m\nabla$ is a bimodule homomorphism from
$\Omega^{1}$ to $\Omega^{2}$. That is, it verifies
$T(f \omega)=fT(\omega)$ and $T(\omega f)=T(\omega)f$.
Torsion is defined by
$$T^{a}\equiv T (b^{a})=db^{a}+m\Gamma^{a}\eqn\tor$$.

The covariant derivative may be extended
as a linear map from the tensor product over ${\cal C}(M)$
of $s$ copies of $\Omega^{1}$ to the tensor product
of $s+1$ copies of $\Omega^{1}$. This is
done recurrently with the following extension of the Leibnitz rule:
$$\nabla(\omega\otimes\omega')=\nabla(\omega)\otimes\omega'+
\sigma_{s}(\omega\otimes\nabla\omega'),\eqn\ext$$
here $\omega$ is in $\Omega^{1}$,
 $\omega'$ is in $\otimes^{s-1}\Omega^{1}$ and
$\sigma_{s}$ is given by
$$\sigma_{s}=\sigma\otimes\overbrace{1\otimes 1\dots\otimes 1}
^{ \rm s-1 \  times}.\eqn\permu$$

A more familiar way of writing this Leibnitz
rule is provided by the covariant derivative along
the direction $X$,
$$\nabla_{X}=\iota_{X}\nabla.\eqn\code$$
The Leibnitz rule \ext  \ reads
$$\nabla_{X}(\omega\otimes\omega')=\nabla_{X}\omega\otimes\omega'
+\omega\nabla_{X}\omega'\eqn\leid.$$
The reason we used equation \ext \ is that it is this one that
will give rise to the non-commutative formulation.

Another extension of the covariant
derivative as a linear map from $\Omega^{*}
\otimes_{{\cal C}(M)}\Omega^{1}$
to itself, $D$,  is given with the aid of the Leibnitz
rule:
$$D(\nu\otimes\omega)=d\nu\otimes\omega+(-1)^{\rm deg \nu}
\nu\cdot\nabla\omega,\eqn\leif$$
where the product $\cdot$ is defined by
$$\nu\cdot(\omega\otimes\omega')=(\nu \wedge\omega)\otimes\omega'.$$
Note that,
when acting on $\Omega^{1}\otimes_{{\cal C}(M)}\Omega^{1}$,
$D$ and $\nabla$  are related
by:
$$D=m\otimes 1\ \nabla+T\otimes 1.\eqn\rel$$
The operator $D^{2}$ is a module homomorphism:
it verifies $D^{2}(f\omega)=fD^{2}\omega$,
where f is an arbitrary  function
and $\omega$ is an arbitrary 1-form. Curvature may
be defined by the equation:
$$D^{2} b^{a}=R^{a}{}_{b}\otimes b^{b}.\eqn\cur$$

Suppose $M$ equipped with a metric and
let $\theta^{a}$ be a moving basis.
The connection form $\omega^{a}{}_{b}$ is defined by the
equation $\nabla \theta^{a}=-\omega^{a}{}_{b}\otimes\theta^{b}$.
The compatibility of the connection with the metric, $g$,
is translated by the equation $\nabla g=0$ which
imposes the antisymmetry of $\omega^{a}{}_{b}$ in the indices
$a$ and $b$. Cartan's structure equations
are given by \tor \  and \cur,
with $b^{a}$ replaced by $\theta^{a}$.

\chapter{Linear connections in non-commutative geometry}

In this section we construct linear connections
on non-commutative algebras.
We first begin in the setting of the differential calculus of
Dubois-Violette and then we propose a generalisation to other
frameworks.

\section{Covariant derivative in derivations based differential calculi}
In this section we use the notation  $\Omega^{p}$ for
$\underline{\Omega}^{p}_{{\cal D}\rm er}$.

A covariant derivative is defined as
a linear map from $\Omega^{1}$ to
$\Omega^{1}\otimes_{{\cal A}}\Omega^{1}$
satisfying the two Leibnitz rules:
$$\nabla(f\omega)= df\otimes\omega+f\nabla\omega,\eqn\leiba$$
$$\nabla(\omega f)=\tau(\omega\otimes df)+\nabla\omega f,\eqn\leibb$$
here $f$ is an element of $\cal A$, $\omega$ is a 1-form. It remains
to define $\tau$ which generalises the permutation
$\sigma$ of commutative geometry. When calculating
$\nabla (f\omega g)$ in two different ways one discovers
that $\tau$ must verify
$$\tau(f\omega\otimes\omega')=f\tau(\omega\otimes
\omega'),\eqn\req$$
so this rules out the permutation \perm \ as a possible candidate
for $\tau$. Another  way of expressing
the permutation \perm \ is provided by the canonical imbedding, $i$ of
$\Omega^{2}$ in $\Omega^{1}\otimes_{{\cal C}(M)}\Omega^{1}$ and
the multiplication map $m$:
$$\sigma=1-2i\circ m.\eqn\si$$
The imbedding $i$ admits a generalisation to the non-commutative
differential calculus of Dubois-Violette. It is given by
$$i(\omega\wedge\omega')(X_{1},X_{2})={1 \over 2}
\Big(\omega(X_{1})\omega'(X_{2})-\omega(X_{2})\omega'(X_{1})\Big)
\eqn\produ.$$
 We shall note $m$ the multiplication map
from $\Omega^{1}\otimes_{{\cal A}}\Omega^{1}$
to  $\Omega^{2}$ given by
$$m(\omega\otimes\omega')=\omega\wedge\omega'.\eqn\pr$$
Note that the canonical imbedding, $i$, is such
that the diagram
$$\matrix{
&\Omega^{1}\otimes_{\cal A}\Omega^{1}
\buildrel m \over \longrightarrow &\Omega^{2} \cr
 &\downarrow m &\downarrow i \cr
&\Omega^{2}\buildrel m \over \longleftarrow & \Omega^{1}\otimes_{\cal A}
\Omega^{1}
}\eqn\dia$$
is commutatif.
Note also that $i\circ m( f\omega\otimes\omega')=
fi\circ m(\omega\otimes\omega')$.
So we propose for $\tau$ the following natural expression:
$$\tau=1-2i\circ m.\eqn\to$$
When acting on derivations
we have
$$\tau(\omega\otimes\omega')(X_{1},X_{2})=\omega(X_{2})\omega'(X_{1}).
\eqn\ta$$
This ends up the definition of the covariant
derivative for the  non-commutative differential
calculus based on derivations. Contrary
to the commutative case,
the two equations
\leiba \ and \leibb \ with $\tau$ given by \to \
are not equivalent.

Let $b^{a}$ be a set of 1-forms generating $\Omega^{1}$
as an $\cal A$-bimodule. That is a minimal set of 1-forms such that
an arbitrary 1-form
may be written as
a sum of terms of the form $f b^{a} g$ with $f$ and $g$
elements of $\cal A$. The covariant derivative
of a 1-form is uniquely determined by
$\nabla b^{a}=-\Gamma^{a}$ which defines the linear connection.
This is so because using equations \leiba \ and \leibb \
one gets:
$$\nabla (f b^{a} g)= df\otimes b^{a}g-f \Gamma^{a} g
+f\tau(b^{a}\otimes dg).\eqn\der$$
Note that in formulations using only the left-module
structure of $\Omega^{1}$, that is postulating only equation \leiba,
the covariant derivative is, in general, determined by a
bigger set of forms: one needs a set generating $\Omega^{1}$
as an $\cal A$-left module. Note also that contrary to the
commutative case, a connection 1-form cannot, in general,
be defined since $\Gamma^{a}$ cannot be expressed
as $\omega^{a}{}_{b}\otimes b^{b}$.

Thanks to our two Leibnitz rules the extension
of the covariant derivative may be immediately
done to the tensor product over $\cal A$ of $s$ copies of
$\Omega^{1}$. This is accomplished by the rule:
$$\nabla(\omega\otimes\omega')=\nabla\omega\otimes\omega'
+\tau_{s}(\omega\otimes\nabla\omega'),\eqn\exte$$
where $\omega$ is in $\Omega^{1}$, $\omega'$
in $\otimes^{s-1}_{\cal A}\Omega^{1}$ and $\tau_{s}$ is given by
a relation analogous to \permu:
$$\tau_{s}=\tau\otimes\overbrace{1\otimes 1 \otimes \dots\otimes1}
^{\rm s-1 \ times}.\eqn\permut$$

In order to make the Leibnitz rules
\leiba \ and \leibb \ more transparent one can define
the covariant derivative along a derivation $X$ as in \code. The
Leibnitz rules \leiba \ and \leiba, for this case, yield:
$$\nabla_{X}(f\omega)=X(f)\omega+f\nabla_{X}\omega,\eqn\derx$$
$$\nabla_{X}(\omega f)=\omega X(f)+\nabla_{X}\omega f.\eqn\dery$$
These formulae are a natural requirement for the covariant
derivative in the context
of the differential calculus of Dubois-Violette.
Similarly, the Leibnitz rule \exte \
may be made more transparent in this case with the aid of $\nabla_{X}$:
$$\nabla_{X}(\omega\otimes\omega')=
\nabla_{X}\omega\otimes\omega'+\omega\otimes\nabla_{X}\omega'.\eqn\derz$$

The other extension, $D$, may be defined in a way similar
to the commutative case as a linear map from
$\Omega^{*}\otimes_{\cal A}\Omega^{1}$ to itself verifying the Leibnitz
rule:
$$D(\nu\otimes\omega)=d\nu\otimes\omega+(-1)^{\rm deg\nu}
\nu\cdot\nabla\omega.\eqn\oth$$

Torsion is defined as in the previous section. The
operator $T=d-m\nabla$ verifies $T(f\omega g)=fT(\omega)g$
for all $f$ and $g$ in $\cal A$ and
$\omega$ in $\Omega^{1}$. Torsion is the 2-form defined by:
$$T^{a}=T(b^{a})=db^{a}+m\Gamma^{a}.\eqn\torsi$$

Curvature may be defined with the operator
$D^{2}$ which verifies $D^{2}(f\omega g)=f(D^{2}\omega)g$.
In general $D^{2}b^{a}$ cannot be written in the form
$R^{a}{}_{b}\otimes b^{b}$. So curvature is defined as
the element of $\Omega^{2}\otimes_{\cal A}\Omega^{1}$ given
by
$$R^{a}=D^{2}b^{a}.\eqn\courb$$

\section{Linear connections in formalisms without derivations}

Derivations were important ingredients for the construction
of linear connections in the setting of non-commutative
geometry of Dubois-Violette. Apart from their use in
the construction of $\underline{\Omega}_{{\cal D}er}$, they permitted
us to define the linear map $\tau$ which appears
in the Leibnitz rule using the right-module structure
of the set of 1-forms. The generalisation of
the construction of the previous section to differential
calculi that do not use derivations is thus reduced
to the definition of $\tau$. If we had a natural
imbedding of $\Omega^{2}$ in $\Omega^{1}\otimes_{{\cal A}}
\Omega^{1}$ we could  use the expression
\to \ for $\tau$.

A natural imbedding
exists for the universal differential algebra,
 in this  case  the tensor product $\Omega^{1}\otimes_{\cal A}\Omega^{1}$
is isomorphic to $\Omega^{2}$ so that $i\circ m=1, \ \tau=-1$ and
the covariant derivative coincides with $\delta-T$, $\delta$
is the universal exterior derivative and $T$ is a bimodule homomorphism
defining torsion.

Consider the case of the Connes'differential calculus
with the  Dirac operator. The scalar product \sca \
provides us with a natural imbedding of $\Omega^{2}_{\Di}$
in $\Omega^{1}_{\Di}\otimes_{\cal A}\Omega^{1}_{\Di}$.
 In order to see this
let $P$ be the projector on the orthogonal
of $\rho^{*}(\delta J_{gr}^{1})$. An element $\nu$
of $\Omega^{2}_{\Di}$ may be identified
with an element $\omega=\sum a[\Dir,b][\Dir,c]$
of $\rho^{*}(\Omega^{2})$ verifying $\omega=P\omega$. $\omega$
is uniquely defined. The imbedding $i$ is
thus given by:
$$i(\nu)=\sum a[\Dir,b]\otimes[\Dir,c].\eqn\imbe$$
When $\cal A$ is the commutative algebra of
functions on a compact Euclidien spin manifold and $\Dir$
is the corresponding Dirac operator the above
definition of $\tau$ reduces to that of the permutation $\sigma$.

In other differential calculi, such as those on the quantum plane [\wes]
or on quantum groups [\wor], a natural imbedding may not exist. In
this case, we look for a generalisation of $\tau$ while
keeping some minimal properties.

In the previous cases, $\tau$ was a
bimodule homomorphism from $\Omega^{1}\otimes_{{\cal A}} \Omega^{1}$
to itself.
The consistency of the definition of the covariant
derivative, however, imposes on $\tau$ merely to be a left module
homomorphism. Another requirement is for torsion
defined by \tor \ to be a bimodule homomorphism. This
imposes the condition
$$m\circ\tau=-m.\eqn\cond$$
 So, the minimal requirements one may ask from $\tau$ are
given by \req  \ and \cond.

\section{An example: linear connections on matrix algebras.}

In this section we study the example of
$n \times n$ matrices  to illustrate the constructions of
the
previous section.
The derivations based differential calculus on $M_{n}(C)$
has been studied in [\dubc]. We will use this setting for
the construction of the linear connection.

Derivations on
$M_{n}(C)$ are all inner. A basis of ${\cal D}$er as
a vector space over $C$
is given by:
$$e_{i}=ad_{\lambda_i},\quad i=1,\dots n^{2}-1,\eqn\ba$$
where the $\lambda_{i}$ form a basis of selfadjoint traceless matrices.
A convenient set of 1-forms is provided by the duals of
$e_i$, $\theta^{i}$ defined by
$$\theta^{j}(e_i)=\delta^{j}_{i}.\eqn\dua$$
They generate $\Omega^1$ as a left $M_n$ module.
The $\theta^{i}$ commute with matrices,
$$f\theta^{i}=\theta^{i}f \quad \forall f \in M_n,\eqn\prop$$
because of the defining property \dua.
An arbitrary element of $\Omega^{1}$ may be
written as $\omega=\omega_i\theta^i$, where
$\omega_i$ are $n \times n$ matrices.
The exterior derivative of a matrix, $f$, is given by
$$df=e_if\theta^i=[\lambda_i,f]\theta^i.\eqn\dext$$
 We denote the linear connection $-\nabla \theta^{i}$
by $\Gamma^i$. As $\Gamma^i$ belongs to $\Omega^1 \otimes
_{M_{n}} \Omega^1$, it may be written as
$$\Gamma^i=\Gamma^i{}_{jk}\theta^{j}\otimes\theta^{k},\eqn\lin$$
where the $\Gamma^i{}_{jk}$ are matrices.
Note that when acting on $\theta^i\otimes \theta^j$,
$\tau$ is a permutation:
$$\tau(\theta^i\otimes\theta^j)=\theta^j\otimes\theta^i.\eqn\teau$$
The two Leibnitz rules \leiba \ and
\leibb \ as well
as the property \prop \
allow us to calculate the covariant derivative
of a 1-form, $\omega=\omega_i\theta^i$, in two different ways:
$$\eqalign{
&\nabla\omega=e_j(\omega_i)\theta^j\otimes\theta^i-
\omega_i \Gamma^i{}_{jk}\theta^j \otimes\theta^k,\cr
&\nabla\omega=e_j(\omega_i)\theta^j\otimes\theta^i-
\Gamma^i{}_{jk}\omega_i \theta^j \otimes\theta^k.}$$
In order for these two expressions to coincide
for arbitrary $\omega$ it is necessary and sufficient that the
 $\Gamma^i{}_{jk}$
 belong to the center of $M_n$:
$$\Gamma^i{}_{jk}\in C.\eqn\cond$$
Note that the two Leibnitz rules have restricted
the possible linear connections.
Torsion may be calculated from $T\theta^i$
which is given by
$$T\theta^i=d\theta^i+m\Gamma^i={1 \over 2}\Big(-C^i{}{jk}
+\Gamma^{i}_{[jk]}\Big)\theta^j\wedge\theta^k,\eqn\torss$$
where the $C^i{}_{jk}$ are the structure constants of the derivations
$e_i$.
Similarly, curvature may be calculated from
$D^2\theta^i$ which is given by:
$$D^2\theta^i=({1 \over 2}\Gamma^{i}{}_{jn}C^j_{lm}
-\Gamma^{i}{}_{lk}\Gamma^k{}_{mn})\theta^l\wedge\theta^m \otimes
\theta^n.\eqn\macou$$

\chapter{Conclusion}

We proposed a  definition for linear connections on
non-commutative algebras. The covariant derivative was defined
as a linear map from $\Omega^{1}$ to $\Omega^{1}\otimes_{\cal A}
\Omega^{1}$ verifying two
Leibnitz rules which exploit the bimodule structure
of $\Omega^{1}$. The definition was reduced to that of a
map from $\Omega^1\otimes_{\cal A}\Omega^1$ to itself, $\tau$,
which reduces to a
permutation in the commutative case. A natural definition was
proposed for this map in the context of the derivations
based differential calculus of Dubois-Violette and the
example of matrix geometry was examined.
The Dixmier trace was used in order to
define $\tau$ in the setting proposed by Connes.
It would
be interesting to study more examples
in order to illustrate the construction
we proposed in the framework of the Connes' differential calculus.
We leave this to a future work.

The physical motivation of defining linear connections
on non-commutative algebras is the formulation of
gravitational theories on what might be more appropriate
than manifolds for the  description of the small scale
structure of space-time.
 The definition of what would replace the metric
is still lacking. One may define it, as in commutative geometry,
as an element of $\Omega^1\otimes_{\cal A}\Omega^1$ or as
mulilinear mapping from $\Omega^1\otimes_{\cal A}
\Omega^1$ to $\cal A$ [\sit] satisfying some symmetry requirements.
The two definitions are not, in general, equivalent.
It is not clear whether these two definitions are the
only candidates for a metric. Then one has to
impose a metricity-like condition in order
to have what replaces the Levi-Civit\'a connection.
Finally, it remains to construct a Ricci scalar
from the previous ingredients. A different line
of attack has been proposed in [\kal].

Another direction which is worth further investigations
is the construction of the map $\tau$ for
general non-commutative differential calculi
from some defining properties; the ones we mentioned at
the end of section 4.2 being some examples.

\ack

I am grateful to Dr. J. Madore for many helpful discussions
and to Dr. M. Dubois-Violette for  valuable
 suggestions and comments.

\refout
\end